\documentclass{article}
\usepackage{ijcai19}

\usepackage{times}
\usepackage{soul}
\usepackage{url}
\usepackage[hidelinks]{hyperref}
\usepackage[utf8]{inputenc}
\usepackage[small]{caption}
\usepackage{graphicx}
\usepackage{amsmath}
\usepackage{booktabs}
\usepackage{algorithm}
\usepackage{algorithmic}
\usepackage{listings,multicol}
\usepackage[caption=false]{subfig}

\usepackage{oz}
\usepackage{makeidx}
\usepackage{listings}
\usepackage{amsfonts}
\usepackage{relsize}
\usepackage{urwchancal}
\usepackage{stmaryrd}
\usepackage{ dsfont }

\newenvironment{Proof}{\noindent{\bf Proof:}}{$\QED$}

\newtheorem{innercustomthm}{Theorem}
\newenvironment{customthm}[1]
  {\renewcommand\theinnercustomthm{#1}\innercustomthm}
  {\endinnercustomthm}
  
\newtheorem{thm}{Theorem}
\newtheorem{lem}[thm]{Lemma}
  
\newcommand{\uri}[1]{\ensuremath{{:}{\texttt{#1}}}}
\newcommand{\urilam}{\ensuremath{{:}{\lambda}}}
\newcommand{\var}[1]{\ensuremath{\text{?#1}}}
\newcommand{\lit}[1]{\ensuremath{\text{``#1''}}}
\newcommand{\trip}[3]{\ensuremath{\langle #1,\allowbreak #2,\allowbreak #3 \rangle}}
\newcommand{\critical}{\ensuremath{\texttt{critical}}}
\newcommand{\score}{\ensuremath{\texttt{score}}}

\newcommand{\gpattern}[3]{\ensuremath{[ #1,\allowbreak #2,\allowbreak #3 ]}}

\newcommand{\externalimplementationlink}{\url{https://github.com/paolo7/ap2}}

\newcommand{\Nto}{\ensuremath{t^S}} 
\newcommand{\Ntoc}{\ensuremath{t^I}} 
\newcommand{\NtoPrime}{\ensuremath{t_{\score}}} 
\newcommand{\Nta}{\ensuremath{t_A^I}} 
\newcommand{\Ntao}{\ensuremath{t_A^S}} 
\newcommand{\Nts}{\ensuremath{t_A^{\mathds{S}}}}
\newcommand{\NtAPrime}{\ensuremath{t_A^{\prime}}} 
\newcommand{\NtAoPrime}{\ensuremath{t_A^{S\prime}}}
\newcommand{\Ntc}{\ensuremath{t_C}} 
\newcommand{\NtA}{\ensuremath{t_A}} 
\newcommand{\NtAtwo}{\ensuremath{t_A}} 
\newcommand{\Ntq}{\ensuremath{t_q}} 

\begin{document}

\title{Rule Applicability on RDF Triplestore Schemas}

\author{
Paolo Pareti$^1$
\and
George	Konstantinidis$^1$\and
Timothy J.\ Norman$^1$\And
Murat \c{S}ensoy$^2$
\affiliations
$^1$University of Southampton, Southampton, United Kingdom\\
$^2$\"{O}zye\u{g}in University, Istanbul, Turkey
}

\maketitle

\begin{abstract}

Rule-based systems play a critical role in health and safety, where policies created by experts are usually formalised as rules. When dealing with increasingly large and dynamic sources of data, as in the case of Internet of Things (IoT) applications, it becomes important not only to efficiently apply rules, but also to reason about their applicability on datasets confined by a certain schema. 
In this paper we define the notion of a triplestore schema which models a set of RDF graphs. Given a set of rules and such a schema as input we propose a method to determine rule applicability and produce output schemas. Output schemas model the graphs that would be obtained by running the rules on the graph models of the input schema. We present two approaches: one based on computing a canonical (critical) instance of the schema, and a novel approach based on query rewriting. We provide theoretical, complexity and evaluation results that show the superior efficiency of our rewriting approach.

\end{abstract}

\section{Introduction}

Inference rules are a common tool in many areas where they are used, for example, to model access control policies \cite{Beimel2011OWLandSWRLAccessControl} and business rules \cite{Fortineau2014SWRLapplicationBusinessRules}. In this paper we are motivated by their use in Internet of Things (IoT) applications, where rules are often used to capture human decision making in a simple and straightforward way \cite{Perera2014RulesReview}. This is especially true in safety-critical domains, such as in the field of Occupational Health and Safety (OHS). OHS knowledge is codified by experts into policies, which are then translated into rules to monitor and regulate workplaces. For example, OHS regulations set limits on human exposure to certain gases. Monitoring systems can use these rules to determine, from sensor data, whether dangerous gas concentration levels have been reached, and trigger warnings or perform actions such as increasing ventilation. Real-world use cases have been provided by industrial partners, such as a wireless and electronic solutions company, and OHS policies from the International Labour Organisation \cite{ILO6331}.

An important limitation of current inference rule applications in the IoT domain is that they require expert human interventions not only to create rules, but also to manage them. This includes determining when they are applicable and what types of facts we can ultimately infer. In the gas-concentration example above, human experts would be employed to answer questions such as: could the rule that aggregates sensor data be used, maybe in conjunction with other rules, to determine whether an area should be evacuated? Is this rule applicable to the available data sources? And will this rule still be applicable after the underlying data sources change (e.g., in case sensors stop working or are substituted with others)? Knowing which rules can be applied and what type of facts can be inferred on a dataset can have safety critical implications, as OHS policies might depend on the availability of certain pieces of information. It is important to note that by executing rules on a specific dataset we only discover the facts that are currently entailed. To predict which facts could potentially be inferred in future versions of the dataset, we need to reason about its schema.

As IoT scenarios become increasingly complex and dynamic, managing rules in a timely and cost effective way requires improvements in automation. In this paper we present an approach that can answer these questions automatically, by reasoning about an abstraction of the available data sources, called the \emph{triplestore schema}. We define triplestore schemas as abstract signatures of underlying data, similar to database schemas. Triplestore schemas can be defined by experts or, as we will see later, can be derived from the types of sensor available. Such schemas are \emph{dynamic}, changing as the data sources change; e.g., a new sensor is added to the system creating triples with new predicates, and entailing other facts.

We consider \texttt{RDF} \cite{Hayes2014GeneralisedRDF} triplestores and we model triplestore schemas as sets of \texttt{SPARQL} \cite{2008SPARQL} \emph{triple patterns}, which in some formal sense restrict or model the underlying \texttt{RDF} data. We express rules as \texttt{SPARQL} \texttt{construct} queries. This type of rules model \texttt{SPIN} \cite{knublauch2011spin} inference rules, which correspond to Datalog rules \cite{Ceri1989Datalog}, and are also compatible with the monotonic subsets of other rule languages, such as \texttt{SWRL} \cite{horrocks2004swrl}. Given an input triplestore schema and a set of rules, our objective is to decide whether the rules would apply on some \texttt{RDF} dataset modelled by this schema. We do so by computing the ``output'' or ``consequence'' schema of these hypothetical rule applications: this is the schema that models all possible \texttt{RDF} datasets that can be obtained by executing the rules on all datasets modeled by the input schema. It is worth noting that our approach is only concerned with schemas, and it is compatible with relational datasets, as long as their schema and and rules can be expressed using \texttt{RDF} and \texttt{SPARQL} \cite{Calvanese2017Ontop}.

Reasoning at the schema level has been explored previously for databases \cite{marnette2009generalized} and Description Logics \cite{glimm2014abstraction}. In fact, for a different but closely related problem of reasoning on database schemas (called \emph{chase termination}), Marnette \cite{marnette2009generalized} employed a canonical database instance, called the \emph{critical instance}, which is representative of all database instances of the given schema, on which we base one of our solutions. 

We propose two approaches to reason about the applicability of inference rules on triplestore schemas. First, we re-use the critical instance for our triplestore schemas and develop an approach based on this representative \texttt{RDF} graph: running the rules on this graph produces evaluation mappings which, after careful manipulation in order to account for peculiarities of \texttt{RDF} literals, help produce our consequence schemas. When constructing the critical instance, as in the original case of relational databases, we need to place all constants appearing in our schema and rules in the constructed instance in many possible ways. 
This leads to a blowup in its size and so we turn our attention to finding a much smaller representative \texttt{RDF} graph, that we call the \emph{sandbox} graph, and which we populate with only one ``representative'' element. We then develop a novel query rewriting algorithm that can compute the consequence schema on the sandbox graph. We provide correctness, complexity, and evaluation results and experimentally exhibit the efficiency and scalability of our rewriting-based approach: it surpasses the critical-instance methods by orders of magnitude while scaling to hundreds of rules and schema triples in times ranging from milliseconds to seconds. 

\section{Background} \label{background}

We consider triplestores containing a single \texttt{RDF} \emph{graph}, without blank nodes. Such a graph is a set of \emph{triples} $\mathds{U} \times \mathds{U} \times (\mathds{U} \cup \mathds{L})$ where $\mathds{U}$ is the set of all \texttt{URI}s, $\mathds{L}$ the set of all literals and $\mathds{V}$ the set of all variables. We use the term \emph{constants} to refer to both literals and \texttt{URI}s. A \emph{graph pattern} is a set of \emph{triple patterns} defined in: $(\mathds{U} \cup \mathds{V}) \times (\mathds{U} \cup \mathds{V}) \times (\mathds{U} \cup \mathds{L} \cup \mathds{V})$. 
Given a pattern $P$, $vars(P)$ and $const(P)$ are the sets of variables and constants in the elements of $P$, respectively. We represent \texttt{URI}s as namespace-prefixed strings of characters, where a namespace prefix is a sequence of zero or more characters followed by a column e.g.\ \uri{a}; literals as strings of characters enclosed in double-quotes, e.g.\ \lit{l}, and variables as strings of characters prefixed by a question-mark, e.g.\ \var{v}. The first, second and third elements of a triple $t$ are called, respectively, \emph{subject}, \emph{predicate} and \emph{object}, and are denoted by $t[x]$, $x \in \tau$ with $\tau$ denoting throughout the paper indexes $\tau = \{1,2,3\}$. 

A \emph{variable substitution} is a partial function $\mathds{V} \pfun \mathds{V} \cup \mathds{U} \cup \mathds{L}$.
A \emph{mapping} is a variable substitution defined as $\mathds{V} \pfun \mathds{U} \cup \mathds{L}$. Given a mapping $m$, if $m(?v) = n$, then we say $m$ contains \emph{binding} $?v \to n$. The domain of a mapping $m$ is the set of variables $dom(m)$.  
Given a triple or a graph pattern $p$ and a variable substitution $m$ we abuse notation and denote by $m(p)$ the pattern generated by substituting every occurrence of a variable $?v$ in $p$ with $m(?v)$ if $?v\in dom(m)$ (otherwise $?v$ remains unchanged in $m(p)$). 

Given a graph pattern $P$ and a graph $G$, the \texttt{SPARQL} evaluation of $P$ over $G$, denoted with $\llbracket P\rrbracket_G$, is a set of mappings as defined in \cite{Perez2006Semantics}. 
 A graph pattern \emph{matches} a graph if its evaluation on the graph returns a non-empty set of mappings. 
We consider inference rules $A \rightarrow C$, where $A$ and $C$ are graph patterns, and can be expressed as \texttt{SPARQL} \texttt{construct} queries. Note that essentially both $A$ and $C$ in a rule are conjunctive queries~\cite{abiteboulbook}. The \emph{consequent} $C$ of the rule is represented in the \texttt{construct} clause of the query, which is instantiated using the bindings obtained by evaluating the \emph{antecedent} $A$, expressed in the \texttt{where} clause. A single application of a rule $r: A \rightarrow C$ to a dataset $I$, denoted by $r(I)$, is 
$I \cup \bigcup_{m \in \llbracket A \rrbracket_I} \left \{ m(C), \; \text{if} \; m(C)\;  \text{is a valid \texttt{RDF} a graph} \right \} $. 
Rule notations such as \texttt{SPIN} and \texttt{SWRL} can be represented in this format \cite{Nick2018SWRL2SPIN}. The closure, or saturation, of a dataset $I$ under a set of inference rules $R$, denoted by $clos(I,R)$, is the unique dataset obtained by repeatedly applying all the rules in $R$ until no new statement is inferred, that is,  $clos(I,R) = \bigcup_{i=0}^{i=\infty} I_i$, with  $I_0 = I$, and $I_{i+1} = \bigcup_{r\in R}\{r(I_i)\}$.

\section{Problem Description} \label{sec:probdescr}

To reason about schemas we need a simple language to model, and at the same time restrict, the type of triples that an \texttt{RDF} graph can contain. It should be noted that, despite the similarity in the name, the existing \texttt{RDF} \emph{schema} (\texttt{RDFS}) vocabulary is used to describe ontological properties of the data, but not designed to restrict the type of triples allowed in the dataset. 
In this paper we define a \emph{triplestore schema} (or just \emph{schema}) $S$ as a pair $\langle S^G, S^\Delta \rangle$, where $S^G$ is a set of triple patterns, and $S^\Delta$ is a subset of the variables in $S^G$ which we call the \emph{no-literal} set. Intuitively, $S^G$ defines the type of triples allowed in a database, where variables act as wildcards, which can be instantiated with any constant element. 

To account for the restrictions imposed by the \texttt{RDF} data model, the no-literal set $S^\Delta$ defines which variables cannot be instantiated with literals, thus $S^\Delta$ must at least include all variables that occur in the subject or predicate position in $S^G$. For example, if $\trip{\var{v1}}{\uri{a}}{\var{v2}} \in S^G_{'}$ and $\var{v2} \not \in S^\Delta_{'}$, then the instances of schema $S_{'}$ can contain any triple that has \uri{a} as a predicate. If $\trip{\uri{b}}{\uri{c}}{\var{v3}} \in S^G_{'}$ and $\var{v3} \in S^\Delta_{'}$, the instances of $S_{'}$ can contain any triple that has \uri{b} as a subject, \uri{c} as a predicate, and a \texttt{URI} as an object. To prevent the occurrence of complex interdependencies between variables, we restrict each variable to occur only once (both across triples, and within each triple) in $S^G$ and in the rule consequents. 

A graph $I$ \emph{is an instance of a schema} $S$ if 
for every triple $t^I$ in $I$ there exists a triple pattern $t^S$ in $S^G$, and a mapping $m$ such that (1) $m(t^S) = t^I$ and (2) $m$ does not bind any variable in $S^\Delta$ to a literal. In this case we say that $S$ \emph{models} graph $I$ (and that each triple $t^S$  models triple $t^I$). All instances of $S$ are denoted by $\mathds{I}(S)$. We say that two schemas $S$ and $S'$ are semantically equivalent if they model the same set of instances (formally, if $\mathds{I}(S) = \mathds{I}(S')$). 
Notice that any subset of an instance of a schema is still an instance of that schema. 
A rule $r$ within a set of rules $R$ is \emph{applicable} with respect to a triplestore schema $S$ if there exists a graph $I$ instance of $S$, such that the precondition of $r$ matches $clos(I,R-r)$. 

Consider the following example scenario of a mine where sensors produce data modelled according to the Semantic Sensor Network Ontology (SSN) \cite{Taylor2017SSN}, 
with namespace \texttt{sosa}. 
In SNN, sensor measurements are called \emph{observations}. A simple approach to create the schema of a dataset is the following. A dataset that can be populated with sensor measurements of a property \uri{x} (e.g., temperature) can be defined with triple pattern $\gpattern{\var{v1}}{\texttt{sosa:observedProperty}}{\uri{x}}$. Pattern $\gpattern{\var{v1}}{\texttt{sosa:hasResult}}{\var{v2}}$ indicates that the results of these measurements are collected and pattern $\gpattern{\var{v1}}{\texttt{sosa:hasFeatureOfInterest}}{\uri{y}}$ indicates that the results are applicable to a specific entity \uri{y} (e.g., a room). 
Similar patterns are presented in \cite{Chaochaisit2016CSVX} in the context of converting \texttt{CSV} data into the SNN format . 
In this example, the sensors are deployed only in one tunnel, namely $\uri{TunnelA}$, and schema $S_1$ is:
\scriptsize \begin{lstlisting}
$S^G_1$ = $\{ \gpattern{\var{v1}}{\texttt{sosa:observedProperty}}{\uri{CO\_Danger}}$,
      $\gpattern{\var{v2}}{\texttt{sosa:observedProperty}}{\uri{WorkerTag}}$,
      $\gpattern{\var{v3}}{\texttt{sosa:hasFeatureOfInterest}}{\uri{TunnelA}}$,
      $\gpattern{\var{v5}}{\texttt{sosa:hasResult}}{\var{v4}}\}$
$S^\Delta_1$ = $\{\var{v1}, \var{v2}, \var{v3}, \var{v5}\}$      
\end{lstlisting} \normalsize

We now consider instance $I_1$ of schema $S_1$. In this instance, the sensors in tunnel A observed both a dangerous gas concentration, represented by the value \lit{1}, and the presence of worker \uri{John}.
\scriptsize \begin{lstlisting}
$I_1$ = $\{ \trip{\uri{o1}}{\texttt{sosa:observedProperty}}{\uri{CO\_Danger}}$,
      $\trip{\uri{o1}}{\texttt{sosa:hasFeatureOfInterest}}{\uri{TunnelA}}$,
      $\trip{\uri{o1}}{\texttt{sosa:hasResult}}{``1"}$,
      $\trip{\uri{o2}}{\texttt{sosa:observedProperty}}{\uri{WorkerTag}}$,
      $\trip{\uri{o2}}{\texttt{sosa:hasFeatureOfInterest}}{\uri{TunnelA}}$,
      $\trip{\uri{o2}}{\texttt{sosa:hasResult}}{\uri{John}}\}$
\end{lstlisting} \normalsize

Consider two rules $r_1$ and $r_2$. The first one detects when workers trespass on an ``off-limit'' area, and the second one labels areas with dangerous gas concentrations as ``off-limit''.

\scriptsize \begin{lstlisting}
$r_1$ = $\{ \gpattern{\var{v1}}{\texttt{sosa:observedProperty}}{\uri{WorkerTag}}$,
      $\gpattern{\var{v1}}{\texttt{sosa:hasFeatureOfInterest}}{\var{v2}}$,
      $\gpattern{\var{v1}}{\texttt{sosa:hasResult}}{\var{v3}}$,
      $\gpattern{\var{v2}}{\texttt{rdf:type}}{\uri{OffLimitArea}}\}$
      $\rightarrow$ $\{ \gpattern{\var{v2}}{\texttt{rdf:type}}{\uri{TrespassedArea}}\}$
$r_2$ = $\{ \gpattern{\var{v1}}{\texttt{sosa:observedProperty}}{\uri{CO\_Danger}}$,
      $\gpattern{\var{v1}}{\texttt{sosa:hasFeatureOfInterest}}{\var{v2}}$,
      $\gpattern{\var{v1}}{\texttt{sosa:hasResult}}{``1"}\}$ 
      $\rightarrow$ $\{ \gpattern{\var{v2}}{\texttt{rdf:type}}{\uri{OffLimitArea}}\}$
\end{lstlisting} \normalsize

Since the precondition of rule $r_2$ matches dataset $I_1$, we can apply the rule and derive a new fact: $\gpattern{\uri{TunnelA}}{\texttt{rdf:type}}{\uri{OffLimitArea}}$. On the instance extended by this new fact, rule $r_1$ is applicable and adds $\gpattern{\uri{TunnelA}}{\texttt{rdf:type}}{\uri{TrespassedArea}}$.

Our approach relies on being able to decide which rules are applicable on a specific triple store schema, e.g., $S_1$, in absence of any particular instance, e.g., $I_1$. Since the precondition of rule $r_2$ matches dataset $I_1$, this rule is directly applicable on schema $S_1$, and we would like to be able to decide this by only looking at $S_1$. Moreover if we can decide this and extend schema $S_1$ with a triple pattern that is the schema of  $\gpattern{\uri{TunnelA}}{\texttt{rdf:type}}{\uri{OffLimitArea}}$ (in this case that schema would be the same triple itself), then we would able to reason with this new schema and decide that rule $r_1$ is also applicable.
In practice, what we would like to do is to compute a schema that captures all consequences of applying our set of rules on any potential instance.

The following definition captures this intuition.
Given a schema $S$ and a set of rules $R$, a schema $S^{\prime}$ is a schema consequence of $S$ with respect to $R$, denoted $con(S,R)$, if $\mathds{I}(S^{\prime}) = \bigcup_{I \in \mathds{I}(S)} \{I' | I' \subseteq clos(I,R)$\}. We can notice that since every subset of an instance of a schema is still an instance of that schema, a dataset can contain the consequence of a rule application without containing a set of triples matching the antecedent. This situation is commonly encountered when some triples are deleted after an inference is made.

Keeping track of the schema consequences allows us to directly see which rules are applicable to instances of a schema without running the rules on the data. 
In correspondence to a single rule application $r(I)$, of a rule $r$ on an instance $I$, we define a \emph{basic consequence} of a schema $S$ by a rule $r$, denoted by $r(S)$, as a finite schema $S^{\prime}$ for which $\mathds{I}(S') = \bigcup_{I \in \mathds{I}(S)} \{I' | I' \subseteq r(I)\}$. It is now easy to see that the consequence schema for a set of rules $con(S,R)$ is obtained by repeatedly executing $r(S)$ for all $r \in R$ until no new pattern is inferred. Formally, $con(S,R) = \bigcup_{i=0}^{i=n} S_i$, with  $S_0 = S$, and $S_{i+1} = \bigcup_{r\in R}\{r(S_i)\}$, and $S_n= S_{n-1}$ (modulo variable names).
In the following section we focus on the core of our problem which is computing a single basic schema consequence $r(S)$, and describe two approaches for this, namely Schema Consequence by Critical Instance ($\critical(S,r)$), and Schema Consequence by Query Rewriting ($\score(S,r)$). 

\section{Computing the Basic Schema Consequence}
\label{sec:approach}

Given a schema $S$ and a rule $r: A \to C$, our approach to compute the basic schema consequence for $r$ on $S$ is based on evaluating $A$, or an appropriate rewriting thereof, on a ``canonical'' instance of $S$, representative of all instances modelled by the schema. The mappings generated by this evaluation are then (1) filtered (in order to respect certain literal restrictions in \texttt{RDF})  and (2) applied appropriately to the consequent $C$ to compute the basic schema consequence.

We present two approaches, that use two different canonical instances. The first instance is based on the concept of a \emph{critical instance}, which has been investigated in the area of relational databases before~\cite{marnette2009generalized} (and similar notions in the area of Description Logics~\cite{glimm2014abstraction}). Adapted to our \texttt{RDF} setting, the critical instance would be created by substituting the variables in our schema, in all possible ways, with constants chosen from the constants in $S^G$ and $A$ as well as a new fresh constant not in $S^G$ or $A$. In~\cite{marnette2009generalized} this instance is used in order to decide Chase termination; 
Chase is referred to rule inference with \emph{existential} rules, more expressive than the ones considered here and for which the inference might be infinite (see \cite{benedikt2017benchmarking} for an overview of the Chase algorithm). Although deciding termination of rule inference is slightly different to computing the schema consequence, we show how we can actually take advantage of the critical instance in order to solve our problem. Nevertheless, this approach, that we call \critical, creates prohibitively large instances when compared to the input schema. Thus, later on in this section we present a rewriting-based approach, called \score, that runs a rewriting of the rule on a much smaller canonical instance of the same size as $S^G$.

\noindent {\bf The Critical Approach.} For both versions of our algorithms we will use a new fresh \texttt{URI} $\urilam$ such that $\urilam \not\in const(S^G) \cup const(A)$. 
Formally, the critical instance $\mathds{C}(S,A \rightarrow C)$ is the set of triples: 

\[ \hspace{-0.8cm}\{t \lvert\text{ triple }t\text{ with }t[i] =  \left \{ \begin{tabular}{ll} 
$c$ & if $t^S[i]$ is a variable and: 
\\
 & (1) $c$ is a \texttt{URI} or 
 \\
 & (2) $i=3$ and $t^S[i] \not\in S^\Delta$
\\ 
$t^S[i]$ & if $t^S[i]$ is not a variable
\end{tabular} \right \},\\ t^S \in S^G, i \in \tau, c \in const(S^G) \cup const(A) \cup \{\urilam\} \}\]
The critical instance replaces variables with \texttt{URI}s and literals from the set $const(S^G) \cup const(A) \cup \{\urilam\}$, while making sure that the result is a valid \texttt{RDF} graph (i.e.\ literals appear only in the object position) and that it is an instance of the original schema (i.e.\ not substituting a variable in $S^\Delta$ with a literal). 
In order to compute the triples of our basic schema consequence for rule $r$ we evaluate $A$ on the critical instance, and post-process the mappings $\llbracket A\rrbracket_{\mathds{C}(S,r)}$ as we will explain later. 
Before presenting this post-processing of the mappings we stretch the fact that this approach is inefficient and as our experiments show, non scalable.  For each triple $t$ in the input schema $S$, up to $\lvert const(S^G) \cup const(A) \cup \{\urilam\}\rvert ^{vars(t)}$ new triples might be added to the critical instance.

\noindent{{\bf The Score Approach.}} To tackle the problem above we present a novel alternative solution based on query rewriting, called \score . This alternative solution uses a small instance called the \emph{sandbox} instance which is obtained by taking all triple patterns of our schema graph $S^G$ and substituting all variables with the same fresh \texttt{URI} $\urilam$. This results in an instance with the same number of triples as $S^G$. Formally, a sandbox graph $\mathds{S}(S)$ is the set of triples:
\[ \hspace{-0.8cm}\{t \lvert\text{ triple }t\text{ with }t[i] =  \left \{ \begin{tabular}{ll} 
$\urilam$ & if $t^S[i]$ is a variable,
\\ 
$t^S[i]$ & else
\end{tabular} \right \},\\ t^S \in S^G, i \in \tau \}\]
Contrary to the construction of the critical instance, in our sandbox graph, variables are never substituted with literals (we will deal with \texttt{RDF} literal peculiarities in a post-processing step). Also notice that $\mathds{S}(S) \in \mathds{I}(S)$ and $\mathds{S}(S) \subseteq \mathds{C}(S,r)$. 
As an example, consider the sandbox graph of schema $S_1$ from Section \ref{sec:probdescr}:
\scriptsize \begin{lstlisting}
$\mathds{S}(S_1$) = $\{ \gpattern{\urilam}{\texttt{sosa:observedProperty}}{\uri{CO\_Danger}}$,
      $\gpattern{\urilam}{\texttt{sosa:observedProperty}}{\uri{WorkerTag}}$,
      $\gpattern{\urilam}{\texttt{sosa:hasFeatureOfInterest}}{\uri{TunnelA}}$,
      $\gpattern{\urilam}{\texttt{sosa:hasResult}}{\urilam}\}$
\end{lstlisting} \normalsize
The critical instances $\mathds{C}(S_1,r_1)$ and $\mathds{C}(S_1,r_2)$ from our example would contain all the triples in $\mathds{S}(S_1)$, plus any other triple obtained by substituting some variables with constants other than $\urilam$, such as the triple: $\gpattern{\urilam}{\texttt{sosa:hasResult}}{\uri{OffLimitArea}}\}$. A complete example of $\mathds{C}(S_1,r_2)$ is available in our appendix.

In order to account for all mappings produced when evaluating $A$ on $\mathds{C}(S,r)$ we will need to evaluate a different query on our sandbox instance, essentially by appropriately rewriting $A$ into a new query. 
To compute mappings, we consider a rewriting $\mathds{Q}(A)$ of $A$, which expands each triple pattern $t_A$ in $A$ into the union of the 8 triple patterns that can be generated by substituting any number of elements in $t_A$ with $\urilam$. Formally, $\mathds{Q}(A)$ is the conjunction of disjunctions of triple patterns:
\begin{equation*}
\mathds{Q}(A) = \bigwedge{}_{t \in A} \Bigg( \bigvee{}_{\substack{x_1 \in \{\urilam,t[1]\} \\ x_2 \in \{\urilam,t[2]\} \\ x_3 \in \{\urilam,t[3]\} }} 
\trip{x_1}{x_2}{x_3} \Bigg)
\end{equation*}
When translating this formula to SPARQL we want to select mappings that contain a binding for all the variables in the query, so we explicitly request all of them in the select clause. 
For example, consider graph pattern $A_1 = \{\trip{\var{v3}}{\uri{a}}{\var{v4}}, \trip{\var{v3}}{\uri{b}}{\uri{c}}\}$, which is interpreted as query: 
\scriptsize \begin{lstlisting}
SELECT ?v3 ?v4 WHERE { ?v3 :a ?v4 . ?v3 :b :c }
\end{lstlisting}  \normalsize
Query rewriting $\mathds{Q}(A_1)$ then corresponds to:
\scriptsize \begin{lstlisting}
SELECT ?v3 ?v4 WHERE {
  { {?v3 :a ?v4} UNION {:$\lambda$ :a ?v4} UNION {?v3 :$\lambda$ ?v4}
    UNION {?v3 :a :$\lambda$} UNION {:$\lambda$ :$\lambda$ ?v4} UNION {:$\lambda$ :a :$\lambda$}
    UNION {?v3 :$\lambda$ :$\lambda$} UNION {:$\lambda$ :$\lambda$ :$\lambda$} }
  { {?v3 :b :c} UNION {:$\lambda$ :b :c} UNION {?v3 :$\lambda$ :c}
    UNION {?v3 :b :$\lambda$} UNION {:$\lambda$ :$\lambda$ :c} UNION {:$\lambda$ :b :$\lambda$}
    UNION {?v3 :$\lambda$ :$\lambda$} UNION {:$\lambda$ :$\lambda$ :$\lambda$} } }
\end{lstlisting}  \normalsize

Below we treat $\mathds{Q}(A)$ as a union of conjunctive queries, or \texttt{UCQ} \cite{abiteboulbook}, and denote  $q \in \mathds{Q}(A)$ a conjunctive query within it.

Having defined how the \critical\ and \score\ approaches compute a set of mappings, we now describe the details of the last two phases required to compute a basic schema consequence.

\noindent {\bf Filtering of the mappings} This phase deals with processing the mappings computed by either  \critical\ or \score, namely $\llbracket A\rrbracket_{\mathds{C}(S,r)}$ or $\llbracket\mathds{Q}(A)\rrbracket_{\mathds{S}(S)}$. It should be noted that it is not possible to simply apply the resulting mappings on the consequent of the rule, as such mappings might map a variable in the subject or predicate position to a literal, thus generating an invalid triple pattern. Moreover, it is necessary to determine which variables should be included in the no-literal set of the basic schema consequence. 
The schema $S^{\prime}$, output of our approaches, is initialised with the same graph and no-literal set as $S$ (i.e.\ $S^{\prime G} = S^G$, $S^{\prime \Delta} = S^\Delta$). We then incrementally extend $S^{\prime}$ on a mapping-by-mapping basis until all the mappings have been considered, at which point, $S^{\prime}$ represents the final output of our approach. 

For each mapping $m$ in $\llbracket A\rrbracket_{\mathds{C}(S,r)}$ or $\llbracket\mathds{Q}(A)\rrbracket_{\mathds{S}(S)}$, we do the following. We create a temporary no-literal set $\Delta^m$. This set will be used to keep track of which variables could not be bound to any literals if we evaluated our rule antecedent $A$ on the instances of $S$, or when instantiating the consequence of the rule. We initialise $\Delta^m$ with all the variables of our rule $A\rightarrow C$ that occur in the subject or predicate position in some triple of $A$ or $C$, as we know that they cannot be matched to or instantiated with literals. 

Then, we consider the elements that occur in the object position in the triples $t_A$ of $A$. We take all the rewritings $t_q$ of $t_A$ in $\mathds{Q}(A)$ (if using \critical , it would be enough to consider a single rewriting $t_q$ with $t_q = t_A$). Since the mapping $m$ has been computed over the canonical instance ($\mathds{S}(S)$ or $\mathds{C}(S,r)$ depending on the approach), we know that there exists at least one $t_q$ such that $m(t_q)$ belongs to the canonical instance. 
We compute the set of schema triples $t^S$ that model $m(t_q)$, for any of the above $t_q$. Intuitively, these are the schema triples that enable $t_A$, or one of its rewritings, to match the canonical instance with mapping $m$. If $t_A[3]$ is a literal $l$, or a variable mapped to a literal $l$ by $m$, we check if there exists any $t^S$ from the above such that $t^S[3] = l$ or $t^S[3]$ is a variable that allows literals (not in $S^\Delta$). If such triple pattern doesn't exist, then $m(A)$ cannot be an instance of $S$ since it has a literal in an non-allowed positions, and therefore we filter out or \emph{disregard} $m$. If $t_A[3]$ is a variable mapped to $\urilam$ in $m$, we check whether there exists a $t^S$ such that $t^S[3]$ is a variable that allows literals (not in $S^\Delta$). 
If such $t^S$ cannot be found, we add variable $t_A[3]$ to $\Delta^m$. Intuitively, this models the fact that $t_A[3]$ could not have been bound to literal elements under this mapping. 
Having considered all the triples $t_A \in A$ we filter out mapping $m$ if it binds any variable in $\Delta^m$ to a literal. If $m$ hasn't been filtered out, we say that rule $r$ is applicable, and we use $m$ to expand $S^{\prime}$.

\begin{figure}[!t]
        \center{\subfloat[]{%
        \includegraphics[clip,width=0.9\columnwidth]{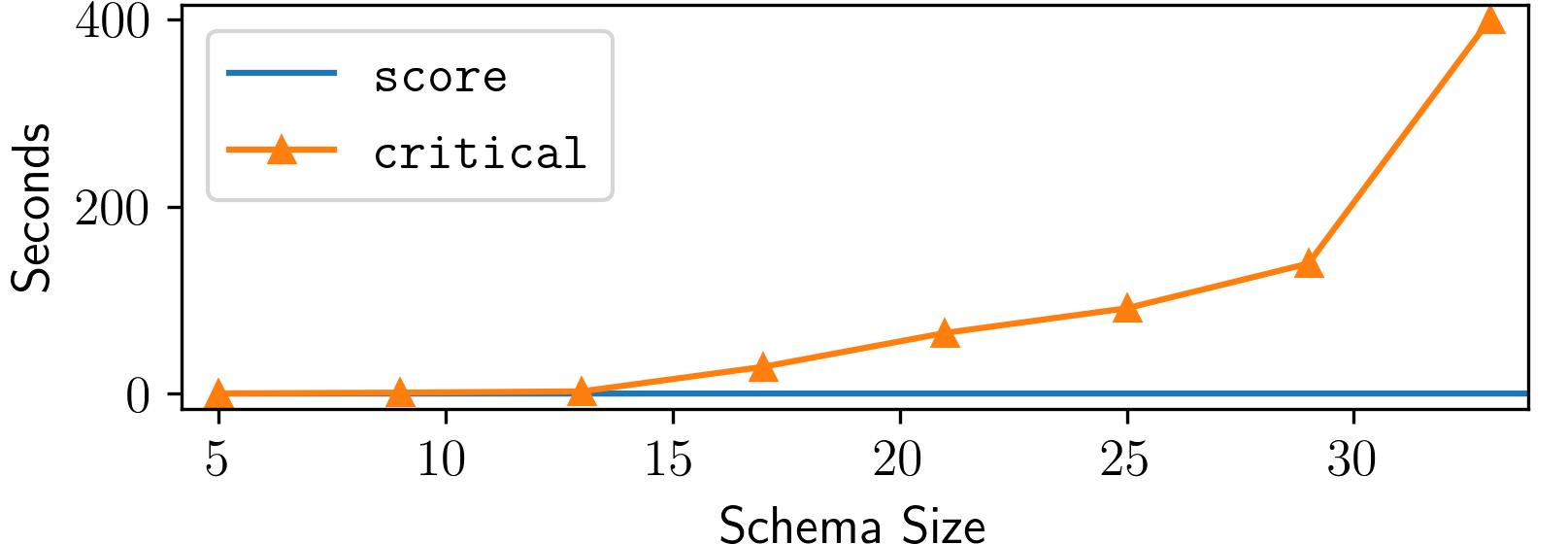}%
        }}\\[-0.1ex]

        \subfloat[]{%
        
        \includegraphics[clip,width=0.9\columnwidth]{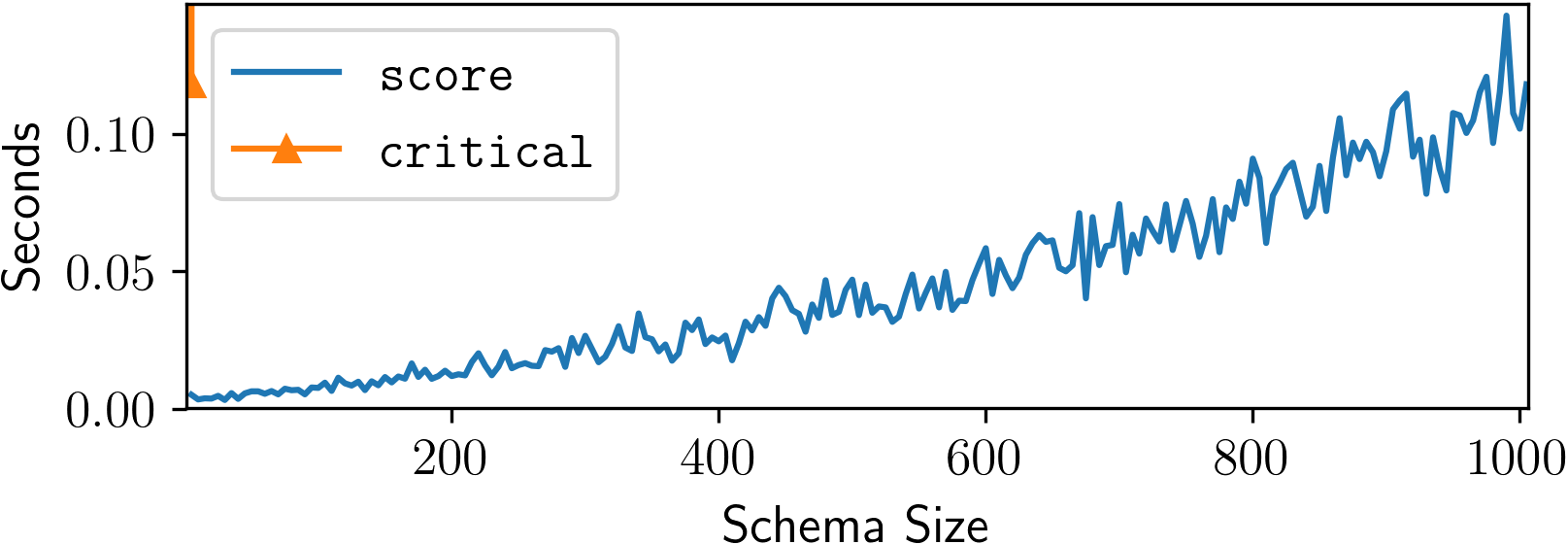}%
        
        }\\[-1.2ex]
        \caption{\label{fig:my-label} Average time to compute 20 schema consequences using \score\ and \critical\ as the schema size $\lvert S\rvert$ grows. 
        The other parameters are: $\lvert P\rvert=1.5\lvert S\rvert$, $\pi_C=0.1$, $\lvert U\rvert =\lvert L\rvert =\lvert S\rvert$, $\lvert R\rvert =4$, $n_A=2$. Due to large difference in performance, subplots (a) and (b) focus, respectively, on \critical\ and \score .}
\end{figure}

\noindent {\bf Schema Expansion.} 
For each mapping $m$ that is not filtered out, we compute the substitution $s^m$, which contains all the bindings in $m$ that map a variable to a value other than $\urilam$, and for every binding $?v \to \urilam$ in $m$, a variable substitution $?v \to ?v^*$ where $?v^*$ is a fresh new variable. We then add triple patterns $s^m(m(C))$ to $S^{\prime G}$ and then add the variables $s^m(\Delta^m) \cap vars(S^{\prime G})$ to $S^{\prime \Delta}$.

Although the schema consequences produced by $\score(S,r)$ and $\critical(S,r)$ might not be identical, they are semantically equivalent. This notion of equivalence is captured by the following theorem.
\begin{customthm}{1} \label{theorem:approachEquality}
  For all rules $r: A \rightarrow C$ and triplestore schemas $S$, $\mathds{I}(\score(S,r)) = \mathds{I}(\critical(S,r))$.
\end{customthm}

The \score\ approach (and by extension also \critical , by Theorem \ref{theorem:approachEquality}) is sound and complete. The following theorem captures this notion by stating the semantic equivalence of $\score(S,r)$ and $r(S)$.
\begin{customthm}{2} \label{newTheoremSimple} For all rules $r: A \rightarrow C$ and triplestore schemas $S$, $\mathds{I}(\score(S,r)) = \mathds{I}(r(S))$.

\end{customthm}
For our proofs, we refer the reader to our appendix.

\noindent{\bf{Termination.}}  It is easy to see that our approaches terminate since our rules do not contain existential variables, and do not generate new \texttt{URI}s or literals (but just fresh variable names). After a finite number of iterations, either approach will only generate isomorphic (and thus equivalent) triple patterns.

\noindent{\bf{Complexity.}} Our central problem in this paper, computing the schema consequence for a set of rules, can be seen as a form of datalog evaluation~\cite{abiteboulbook} on our critical or sandbox instance. Datalog has been extensively studied in databases and its corresponding evaluation decision problem (whether a given tuple is in the answer of a datalog program) is known to be EXPTIME-complete in the so called combined complexity~\cite{vardi82}, and PTIME-complete in data complexity ~\cite{vardi82,dantsin2001complexity}. Data complexity in databases refers to the setting in which the query (or datalog program) is considered fixed and only the data is considered an input to the problem. 
In our setting, data complexity refers to the expectation that the overall number and complexity of the rules remains small. 
It is not hard to see that the corresponding decision problem, stated below, remains PTIME-complete in data complexity. The intuition is that once we construct the critical instance in polynomial time (or alternatively, we grow our set of rules by a polynomial rule rewriting) we have essentially an equivalent problem to datalog evaluation (our rules being the datalog program, and the canonical instance being the database).
\begin{customthm}{3}
Given triple pattern $t_{S^{\prime}}$ 
and schema $S$ as inputs, the problem of deciding whether $t_{S^{\prime}}$ is in the consequence schema of $S$ for a fixed set of rules $R$ is PTIME-complete.
\end{customthm}

\section{Experimental Evaluation}

\begin{figure}[!t]

         \center{\includegraphics[clip,width=\columnwidth]{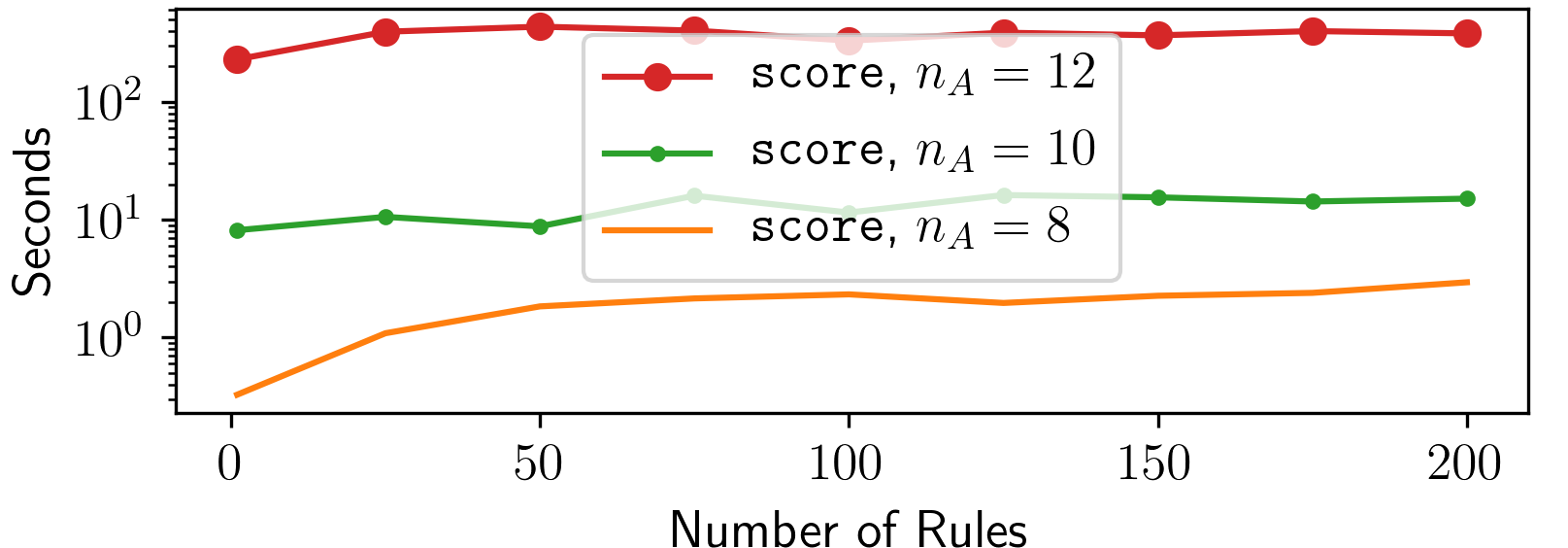}}
        
        \caption{\label{figure:scalabilityScore} Average time to compute 20 schema consequences using \score\ as the number of rules $\lvert R\rvert$ grows, for three configuration of $n_A$. 
        The other parameters are set as follows: $\lvert S\rvert =50$, $\lvert P\rvert=60$, $\pi_C =0.1$, $\lvert U\rvert =\lvert L\rvert =\lvert S\rvert$. }
\end{figure}

We developed a Java implementation of the \score\ and \critical\ approaches and evaluated them on synthetic datasets to compare their scalability. 
We developed a synthetic schema and rule generator that is configurable with 7 parameters: $\pi_C,\lvert P\rvert,\lvert U\rvert,\lvert L\rvert,\lvert S\rvert,\lvert R\rvert, n_A$, 
which we now describe. To reflect the fact that triple predicates are typically defined in vocabularies, our generator does not consider variables in the predicate position. Random triple patterns are created as follows. Predicate \texttt{URI}s are randomly selected from a set of \texttt{URI}s $P$. Elements in the subject and object position are instantiated as constants with probability $\pi_C$, or else as new variables. Constants in the subject positions are instantiated with a random \texttt{URI}, and constants in the object position with a random \texttt{URI} with $50\%$ probability, or otherwise with a random literal. Random \texttt{URI}s and literals are selected, respectively, from sets $U$ and $L$ ($U \cap P = \emptyset$). We consider chain rules where the triples in the antecedent join each other to form a list where the object of a triple is the same as the subject of the next. The consequent of each rule is a triple having the subject of the first triple in the antecedent as a subject, and the object of the last triple as object. An example of such rule generated by our experiment is: 
$\{ \trip{\var{v0}}{\uri{m1}}{\var{v1}}, \trip{\var{v1}}{\uri{m3}}{\var{v2}} \} \rightarrow \{\trip{\var{v0}}{\uri{m2}}{\var{v2}}\} $
In each run of the experiment we populate a schema $S$ and a set of rules $R$ having $n_A$ triples in the antecedent. To ensure that some rules in each set are applicable, half of the schema is initialized with the antecedents triples of randomly selected rules. The other half is populated with random triple patterns. 
We initialize $S^\Delta$ with all the variables in the subject and predicate position in the triples of $S$.  The code used in this experiments is avaliable on GitHub;\footnote{\externalimplementationlink} it uses Apache Jena\footnote{\url{https://jena.apache.org/}} to handle query execution. We run the experiments on a standard Java virtual machine running on an Ubuntu 16.04 computer with 15.5 GB RAM, an Intel Core i7-6700 Processor. Average completion times of over 10 minutes have not been recorded.

Figure \ref{fig:my-label} shows the time to compute the schema consequence for different schema sizes $\lvert S\rvert$ using \critical\ and \score . The parameters have been chosen to be small enough to accommodate for the high computational complexity of the \critical\ approach.  This figure shows that \score\ is orders of magnitude faster, especially on large schema sizes. The \critical\ approach, instead, times out for schema sizes of over 33 triples. 

Figure \ref{figure:scalabilityScore} shows the time to compute the schema consequence for different antecedent sizes $n_A$ and rule numbers $\lvert R\rvert$. The \critical\ approach is not present in this figure, as it timed out in all the configurations. As this figure shows, the \score\ approach can easily scale to a large set of rules.
Given the complexity of \texttt{SPARQL} query answering \cite{Perez2006Semantics}, we can also notice an exponential increase in computation time as more triples are added to the antecedent of a rule. In our experiment setup, the \score\ approach scales to rules with antecedent sizes of up to 12 triples, before timing out.

\section{Related Work} \label{sec:relatedwork}

To the best of the authors' knowledge, our approach is the first to determine the applicability of inference rules to types of \texttt{RDF} triplestores specified by their schema, and to expand their schema with the potential consequences of such rules. Unlike related work on provenance paths for query inferences  \cite{GAD2017}, we do not explicitly model the dependencies between different rules. Instead, we compute their combined potential set of inferences by expanding the original schema on a rule-by-rule basis, through multiple iterations, following the basic principles of the chase algorithm. 
We choose to follow a subset of the \texttt{RDF} data model, and not a simpler graph model such as \emph{generalised} \texttt{RDF} \cite{Hayes2014GeneralisedRDF}, to make our approach more applicable in practice, and compatible with existing tools. We pay particular attention to literals, as they are likely to feature prominently in IoT sensor measurements.

A possible application of our approach is to facilitate the sharing and reusing of inference rules. A relevant initiative in the IoT domain is Sensor-based Linked Open Rules (\texttt{S-LOR}) \cite{Gyrard2017linkedOpenRules}, which provides a comprehensive set of tools to deal with rules, including a rule discovery mechanism. By classifying rules according to sensor types, domain experts can discover and choose which inference rules are most relevant in a given scenario. 
Our approach could automate parts of this process, by selecting rules applicable to the available data sources. 
We refer to \cite{serrano2016review} for a comprehensive review of rule-based reasoning systems applicable to IoT.

Our approach to define a triplestore schema is related to a number of similar languages, and in particular to 
Shape Expressions (ShEx) \cite{Prudhommeaux2014ShapeExpressions} and the Shapes Constraint Language (SHACL) \cite{2017SHACL}. The term \emph{shape}, in this case, refers to a particular constraint on the structure of an \texttt{RDF} graph.  ShEx and SHACL can be seen as very expressive schema languages, and computing schema consequences using such schemas would be impractical. In fact, each inference step would need to consider complex interpendencies between shapes and the values allowed in each triple element, and thus we would generate increasingly larger sets of contrainsts. The triplestore schema proposed in this paper is a simpler schema language and, if we disallow variables in the predicate position, can be modelled as a subset of both ShEx and SHACL.
 
\section{Conclusion}

As its main contribution, this paper presented two approaches to determine the applicability of a set rules with respect to a database schema (i.e.\ if the rule could ever match on any dataset modelled by the schema), by expanding such schema to model the potential facts that can be inferred using those rules, which we call \emph{schema consequence}. 
This can be applied in IoT scenarios, where inference rules are used to aggregate sensor readings from diverse data sources in order to automate health and safety policies.  As the underlying data sources evolve, it is important to determine whether rules are still applicable, and what they can infer.

We focused on \texttt{RDF} triplestores, and on inference rules that can be modelled as \texttt{SPARQL} queries, such as \texttt{SPIN} and \texttt{SWRL}. To do so, we defined a notion of a \emph{triplestore schema} that constrains the type of triples allowed in a graph. This differs from the \texttt{RDF} \emph{schema} (\texttt{RDFS}) specification, which is designed to define vocabularies. While we provide an example on how to describe the schema of simple sensor networks, we think extending this approach to more expressive schema languages could be an interesting venue for future work.

The first of the two approaches that we presented is based on the existing notion of a critical instance; the second on query rewriting. We have theoretically demonstrated the functional equivalence of the approaches, as well as their soundness and completeness. Moreover, we have provided experimental evidence of the superior scalability of the second approach, which can be applied over large schemas and rulesets within seconds.

With this paper we intend to provide a first theoretical framework to reason about rule applicability. We hope that our approach will pave the way, on one hand, for efficient and meaningful policy reasoning in IoT systems, and on the other, for new and interesting rewriting-based schema reasoning approaches in the knowledge representation and databases research areas.

\bibliographystyle{splncs04}
\bibliography{litbib}

\begin{thebibliography}{10}
\providecommand{\url}[1]{\texttt{#1}}
\providecommand{\urlprefix}{URL }
\providecommand{\doi}[1]{https://doi.org/#1}

\bibitem{abiteboulbook}
Abiteboul, S., Hull, R., Vianu, V.: Foundations of databases: the logical
  level. Addison-Wesley Longman Publishing Co., Inc. (1995)

\bibitem{Nick2018SWRL2SPIN}
Bassiliades, N.: {SWRL2SPIN:} {A} tool for transforming {SWRL} rule bases in
  {OWL} ontologies to object-oriented {SPIN} rules. CoRR
  \textbf{abs/1801.09061} (2018), \url{http://arxiv.org/abs/1801.09061}

\bibitem{Beimel2011OWLandSWRLAccessControl}
Beimel, D., Peleg, M.: Editorial: Using owl and swrl to represent and reason
  with situation-based access control policies. Data Knowl. Eng.
  \textbf{70}(6),  596--615 (2011)

\bibitem{benedikt2017benchmarking}
Benedikt, M., Konstantinidis, G., Mecca, G., Motik, B., Papotti, P., Santoro,
  D., Tsamoura, E.: Benchmarking the chase. In: Proceedings of the 36th ACM
  SIGMOD-SIGACT-SIGAI Symposium on Principles of Database Systems. pp. 37--52.
  ACM (2017)

\bibitem{Calvanese2017Ontop}
Calvanese, D., Cogrel, B., Komla{-}Ebri, S., Kontchakov, R., Lanti, D., Rezk,
  M., Rodriguez{-}Muro, M., Xiao, G.: Ontop: Answering {SPARQL} queries over
  relational databases. Semantic Web  \textbf{8}(3),  471--487 (2017)

\bibitem{Ceri1989Datalog}
Ceri, S., Gottlob, G., Tanca, L.: What you always wanted to know about datalog
  (and never dared to ask). IEEE Transactions on Knowledge and Data Engineering
   \textbf{1}(1),  146--166 (1989)

\bibitem{Chaochaisit2016CSVX}
Chaochaisit, W., Sakamura, K., Koshizuka, N., Bessho, M.: {CSV-X: A Linked Data
  Enabled Schema Language, Model, and Processing Engine for Non-Uniform CSV}.
  2016 IEEE Int. Conf. on Internet of Things (iThings) and IEEE Green Computing
  and Communications (GreenCom) and IEEE Cyber, Physical and Social Computing
  (CPSCom) and IEEE Smart Data (SmartData) pp. 795--804 (2016)

\bibitem{Hayes2014GeneralisedRDF}
Cyganiak, R., Wood, D., Markus~Lanthaler, G.: {RDF 1.1 Concepts and Abstract
  Syntax}. {W3C} {R}ecommendation, {W3C} (2014),
  \url{http://www.w3.org/TR/2014/REC-rdf11-concepts-20140225/}

\bibitem{dantsin2001complexity}
Dantsin, E., Eiter, T., Gottlob, G., Voronkov, A.: Complexity and expressive
  power of logic programming. ACM Computing Surveys (CSUR)  \textbf{33}(3),
  374--425 (2001)

\bibitem{Fortineau2014SWRLapplicationBusinessRules}
Fortineau, V., Fiorentini, X., Paviot, T., Louis-Sidney, L., Lamouri, S.:
  Expressing formal rules within ontology-based models using {SWRL}: an
  application to the nuclear industry. International Journal of Product
  Lifecycle Management  \textbf{7}(1),  75--93 (2014), pMID: 65458

\bibitem{glimm2014abstraction}
Glimm, B., Kazakov, Y., Liebig, T., Tran, T.K., Vialard, V.: Abstraction
  refinement for ontology materialization. In: International Semantic Web
  Conference. pp. 180--195. Springer (2014)

\bibitem{Gyrard2017linkedOpenRules}
Gyrard, A., Serrano, M., Jares, J.B., Datta, S.K., Ali, M.I.: {Sensor-based
  Linked Open Rules (S-LOR): An Automated Rule Discovery Approach for IoT
  Applications and Its Use in Smart Cities}. In: 26th International Conference
  on World Wide Web Companion. pp. 1153--1159. WWW '17 (2017)

\bibitem{2008SPARQL}
Harris, S., Seaborne, A.: {SPARQL 1.1 Query Language}. {W3C} {R}ecommendation,
  {W3C} (2013), \url{https://www.w3.org/TR/sparql11-query/}

\bibitem{GAD2017}
Hecham, A., Bisquert, P., Croitoru, M.: {On the Chase for All Provenance Paths
  with Existential Rules}. In: Rules and Reasoning. pp. 135--150. Springer
  International Publishing (2017)

\bibitem{horrocks2004swrl}
Horrocks, I., Patel-Schneider, P.F., Boley, H., Tabet, S., Grosof, B., Dean,
  M., et~al.: {SWRL: A semantic web rule language combining OWL and RuleML}.
  {W3C Member Submission}, {W3C} (2004), https://www.w3.org/Submission/SWRL/

\bibitem{ILO6331}
International Labour Organization: {Act No. 6331 on Occupational Health and
  Safety} (2012),
  \url{https://www.ilo.org/dyn/natlex/natlex4.detail?p_lang=fr&p_isn=92011}

\bibitem{knublauch2011spin}
Knublauch, H.: {SPIN - SPARQL Syntax}. {W3C Member Submission}, {W3C} (2011),
  http://www.w3.org/Submission/spin-sparql/

\bibitem{2017SHACL}
Knublauch, H., Kontokostas, D.: {Shapes constraint language (SHACL)}. {W3C}
  {R}ecommendation, {W3C} (2017), \url{https://www.w3.org/TR/shacl/}

\bibitem{Taylor2017SSN}
Lefran\c{c}ois, M., Cox, S., Taylor, K., Haller, A., Janowicz, K., Phuoc, D.L.:
  {Semantic Sensor Network Ontology}. {W3C} {R}ecommendation, W3C (2017),
  \url{https://www.w3.org/TR/2017/REC-vocab-ssn-20171019/}

\bibitem{marnette2009generalized}
Marnette, B.: Generalized schema-mappings: from termination to tractability.
  In: Proc. of the twenty-eighth ACM SIGMOD-SIGACT-SIGART symp. on Principles
  of database systems. pp. 13--22. ACM (2009)

\bibitem{Perera2014RulesReview}
Perera, C., Zaslavsky, A., Christen, P., Georgakopoulos, D.: {Context Aware
  Computing for The Internet of Things: A Survey}. IEEE Communications Surveys
  Tutorials  \textbf{16}(1),  414--454 (2014)

\bibitem{Perez2006Semantics}
P{\'e}rez, J., Arenas, M., Gutierrez, C.: {Semantics and Complexity of SPARQL}.
  ACM Transactions on Database Systems  \textbf{34}(3),  16:1--16:45 (2009)

\bibitem{Prudhommeaux2014ShapeExpressions}
Prud'hommeaux, E., Labra~Gayo, J.E., Solbrig, H.: {Shape Expressions: An RDF
  Validation and Transformation Language}. In: Proceedings of the 10th
  International Conference on Semantic Systems. pp. 32--40. SEM '14, ACM (2014)

\bibitem{serrano2016review}
Serrano, M., Gyrard, A.: {A Review of Tools for IoT Semantics and Data
  Streaming Analytics}. Building Blocks for IoT Analytics p.~139 (2016)

\bibitem{vardi82}
Vardi, M.Y.: {The Complexity of Relational Query Languages (Extended
  Abstract)}. In: Proceedings of the Fourteenth Annual ACM Symposium on Theory
  of Computing. pp. 137--146. STOC '82, ACM, New York, NY, USA (1982)

\end{thebibliography}

\appendix

\section{Proof of Theorem \ref{theorem:approachEquality}}

\begin{customthm}{1} \label{theorem:approachEquality}
  For all rules $r: A \rightarrow C$ and triplestore schemas $S$, $\mathds{I}(\score(S,r)) = \mathds{I}(\critical(S,r))$.
\end{customthm}

\begin{Proof} Since the mappings generated by the \critical\ and \score\ approaches (namely $\llbracket A\rrbracket_{\mathds{C}(S,r)}$ and $\llbracket \mathds{Q}(A)\rrbracket_{\mathds{S}(S)}$) are post-processed in the same way, we demonstrate the equivalence of the two approaches by demonstrating that ($\Leftarrow$) the set of mappings generated by $\score(S,r)$, which are not filtered out by our post-processing of the mappings, is a subset of the mappings generated by $\critical(S,r)$ and that ($\Rightarrow$) every mapping $\critical(S,r) \setminus \score(S,r)$ is redundant with respect to another mapping in $\critical(S,r) \cap \score(S,r)$. We denote with $\Omega(m,A,S)$ the set of mappings that are not filtered out by our filtering function with respect to $A$ and $S$. We say that a mapping $m$ is redundant with respect of a set of mappings $M$ (and a schema $S$ and rule $r$) if the schema consequences computed over $M$ and $' \setminus \{m\}$ are semantically equivalent.

Note that both the critical and the sandbox instance are constructed by taking a triple in schema $S^G$ and substituting its variables. Thus we can associate each triple in the critical or the sandbox instance with the triples in $S$ that can be used to construct them, and we will call the latter the \emph{origin} triples of the former. For each triple in the critical or the sandbox instance, at least one origin triple must exist. 

$\Rightarrow$) Let a mapping $m \in \llbracket A\rrbracket_{\mathds{C}(S,r)} \setminus \llbracket \mathds{Q}(A)\rrbracket_{\mathds{S}(S)}$. We get all triples $t_A$ of $A$ and for each one we will construct a triple $t_q$ of a conjunctive query $q \in \mathds{Q}(A)$, and mapping $m^{*}$ of $q$ into $\mathds{S}(S)$, such that either $m = m^{*}$ or $m^{*}$ makes $m$ redundant. 
For $i \in \tau$, if $t_A[i]$ is a variable $?v$ and $m(?v)$ $=$ $\urilam$ then set $t_q[i] = ?v$, since there must be a triple in $\mathds{S}(S)$ with $\urilam$ in the $i^{th}$ position so if $m^{*}$ mapped $t_q$ on that triple, $m^{*}(t_q[i]) = m(t_A[i]) = \urilam$ (for all triples of the critical instance that have $\urilam$ in position $i$, its origin triples have variables in the same position, so the sandbox instance would also have $\urilam$ in position $i$). If $t_A[i]$ is a variable $?v$ and $m(?v)$ is a constant $c$ other than \urilam, then we distinguish two cases (a) and (b). 
Let $t_S \in S$ be an origin triple of $m(t_A)$ in the critical instance, then (a) if $t_S[i]=c$ then $\mathds{S}(S)$ would have retained $c$ in the corresponding position in the triple of which $t_S$ is the origin, and so we set $t_q[i]$ to $?v$, in order for $(?v$ $\rightarrow$ $c)$ $\in$ $m$ to also belong to mapping $m^{*}$ of $t_q$ into $\mathds{S}(S)$, or 
(b) if $t_S[i]$ is a variable (we know that $\urilam$ is the element in position $i$ of the triple in the sandbox graph of which $t_S$ is an origin) we consider two sub-cases ($b^{\prime}$) and ($\neg b^{\prime}$). 

We set $t_q[i]$ to $\urilam$ in case ($b^{\prime}$), namely if there is a position $j$ in a triple $t_A^{\prime} \in A$ (different position to $i$, i.e., $j\neq i$ if $t_A = t_A^{\prime}$), for which $t_A^{\prime}[j] = t_A[i]$ and for $t_S^{\prime}$, an origin triple of $t_A^{\prime}$, $t_S^{\prime}[j] = c$. Condition ($b^{\prime}$) essentially says that $?v$ will have to be mapped to $c$ in the sandbox graph due to some other position of the rewriting, so $t_q[i]$ can just be set to $\urilam$ to simply match the corresponding triple in $\mathds{S}(S)$. In case ($\neg b^{\prime}$) we set $t_q[i]$ to $?v$; this condition produces a more \emph{general} mapping $m^{*}$ since $m(?v) = c$ while $m^{*}(?v)$ will be $\urilam$. We say that $m_2$ is more general than $m_1$, denoted by $m_1 \leq_{\lambda} m_2$, if $dom(m_1)$ $=$ $dom(m_2)$ and for all $?v \in dom(m_1)$, either $m_2(?v) = m_1(?v)$ or $m_2(?v)$ $=$ $\urilam$.

Lastly we consider the case of $t_A[i]$ being a constant. If there is an origin triple $t_S$ of $m(t_A)$ in the critical instance such that $t_S[i]$ is the same constant  we also set it to $t_q[i]$. Otherwise, if $t_S[i]$ is a variable in any origin triple $t_S$ we set $t_q[i]$ to $\urilam$. This does not change the mapping under construction $m^{*}$. By following the described process for all triple in $A$ we end up with a rewriting $q$ of $A$ and a mapping $m^{*}$ from $q$ to $ \mathds{S}(S)$ such that $m^{*}$ is more general than $m$.

The existance of the more general mapping $m^{*}$ makes mapping $m$ redundant. This is true because if $m$ is not filtered out by our post-processing, then $m^{*}$ would also not be filtered out (since mappings can only be filtered out only when a variable is mapped to a literal, but $m^{*}$ does not contain any such mapping not already in $m$). If $m$ is not filtered out, then this would lead to the schema being expanded with triples $m(C)$. The instances of the schema obtained through mapping $m$ are a subset of those obtained through mapping $m^{*}$. This can be demonstrated by noticing that for every triple in $m(C)$ there is a corresponding triple in $m^{*}(C)$ that has at each position, either the same constant $c$ or $\urilam$, which is then substituted with a variable in the schema consequence. This variable can always be substituted with the same constant $c$, when instantiating the schema, and therefore there cannot be an instance of the schema consequence generated from mapping $m$ that is not an instance of the schema consequence generated from mapping $m^{*}$. In fact, the only case in which a variable $?v$ in a schema cannot be instantiated with a constant $c$ is when $c$ is a literal and $?v$ is in the no-literal set. However, for all mappings $m \in \llbracket A\rrbracket_{\mathds{C}(S,r)} \setminus \llbracket \mathds{Q}(A)\rrbracket_{\mathds{S}(S)}$, we can notice that whenever our post-processing function adds a variable to the no-literal set, it would have rejected the mapping if that variable was mapped to a literal instead.

$\Leftarrow$) Let a mapping $m^{*} \in \llbracket \mathds{Q}(A)\rrbracket_{\mathds{S}(S)}$. This means that there is a $q \in \mathds{Q}(A)$, for which $m^{*} \in \llbracket q\rrbracket_{\mathds{S}(S)}$. We get all triples $t_q$ of $q$ and for each one we will show that for the triple $t_A$ in $A$, that was rewritten into $t_q$, it holds that either $m^{*}$ is filtered out by the filtering function $\Omega$, or it is also a mapping from $t_A$ into the critical instance (i.e.\ $m^*(t_A) \in \mathds{C}(S,r)$), hence $m^{*} \in \llbracket A\rrbracket_{\mathds{C}(S,r)}$. 
This would prove that the mappings generated by $\llbracket \mathds{Q}(A)\rrbracket_{\mathds{S}(S)}$ are either subsumed by the mappings in $\llbracket A\rrbracket_{\mathds{C}(S,r)}$, or they would be filtered out anyway, and thus $\Omega(\llbracket \mathds{Q}(A)\rrbracket_{\mathds{S}(S)},A,S) \subseteq \Omega(\llbracket A\rrbracket_{\mathds{C}(S,r)},A,S)$. 

For all $t_q \in q$, for all  $i \in \tau$, if $t_q[i]$ is a variable $?v$ then this position has not been changed from rewriting $t_A$ into $t_q$, thus $?v$ will exist in $t_A[i]$; 
if $m^{*}(?v)$ is $\urilam$, then $m^{*}(t_q)[i] = \urilam$, and therefore in any origin triple $t_S$ of $m^{*}(t_q)$, $t_S[i]$ had to be a variable and so $\urilam$ will exist in position $i$ in the triples the critical instance will create for $t_S$, and for these triples $m^{*}$ will partially map $t_A$ onto them. 
Similarly, if $t_q[i]$ is a variable $?v$ and $m^{*}(?v)$ is a constant $c$, then this constant would be present in $t_S[i]$ and so also in all triples coming from $t_S$ in the critical instance; again $m^{*}$ will partially map $t_A$ onto the triples in the critical instance generated from $t_S$. 

If $t_q[i]$ is a constant other than \urilam , then the triple $t_A$ that got rewritten to triple $t_q$ would have the same constant in position $i$. Also, this constant must be present in position $i$ in the triple $m^{*}(t_q)$ of the sandbox graph, and therefore in position $i$ of any of its origin triples $t_S$. In fact, by virtue of how the sandbox graph is constructed, any triple of the schema that does not have this constant in position $i$, cannot be an origin triple of $m^{*}(t_q)$. 
Thus again, $m^{*}$ will partially map $t_A$ onto the triples in the critical instance of which $t_S$ is the origin triple. 

Lastly, if $t_q[i]$ is $\urilam$, then  $m^{*}(t_q)[i] = \urilam$, and in any origin triple $t_S$ of the sandbox triple $m^{*}(t_q)$ we have a variable in position $i$, and in the critical instance we have triples, of which $t_S$ is the origin, with all possible \texttt{URI}s or $\urilam$ in this position; this means that if $m^*(t_A[i])$ is a \texttt{URI}, in the triple $t_A$ that got rewritten in $t_q$, we can ``match'' it in the critical instance: 
if $t_A[i]$ is a variable $?v_1$ then $t_A$ will match the triple in the critical instance (instance of any origin triple $t_S$ of $m^{*}(t_q)$) that contains $m^{*}(?v_1)$ in position $i$ (notice that $m^{*}$ should contain a value for $v_1$ since all variables are present in all rewritings of $A$); else, if $t_A[i]$ is a \texttt{URI} there will be a corresponding image triple (again instance of any $t_S$)  in the critical instance with the same constant in this position. 

If, however, $m^*(t_A[i])$ is a literal, it is possible that there is no triple in the critical instance that can match the literal $m^*(t_A[i])$ in position $i$. We will show that, in this case, the filter function $\Omega$ will remove this mapping. If literal $m^*(t_A[i])$ does not occur in position $i$ in any of the triples in the critical instance that have the same origin triple as $m^*(t_q)$ then, by the definition of how the critical instance is constructed, this must be caused by the fact that either (a) $i \in \{1,2\}$ or (b) for all the possible origin triples $t_S$ of $m^*(t_q)$, $t_S[i] \in S^\Delta$ or $t_S[i]$ is a constant other than $m^*(t_A[i])$. In both cases, our filtering function will filter out the mapping. In fact, it filters out mappings where variables are mapped to literals, if they occur in the subject or predicate position of a triple in $A$. This is true in case (a). Moreover it filters out mappings if a variable in the object position there is a literal $l$, or a variable mapped to a literal $l$, such that there is no origin triple $t_S$ of $m^*(t_q)$ such that $t_S[i] = l$ or $t_S[i]$ is a variable not in $S^{\Delta}$. This is true in case (b). Thus $m^{*}$ is a mapping from $t_A$ into the critical instance. \end{Proof}

\section{Proof of Theorem \ref{newTheoremSimple}}

\begin{customthm}{2} \label{newTheoremSimple} For all rules $r: A \rightarrow C$ and triplestore schemas $S$, $\mathds{I}(\score(S,r)) = \mathds{I}(r(S))$.

\end{customthm}

Theorem \ref{newTheoremSimple} states that $\score(S,r)$ is semantically equivalent to $r(S)$. We prove this by showing that ($\Rightarrow$) every triple that can be inferred by applying rule $r$ on an instance of $S$ is an instance of the schema $\score(S,r) \setminus S$ and that ($\Leftarrow$) every triple in every instance of schema $\score(S,r) \setminus S$ can be obtained by applying rule $r$ on an instance of $S$. To reduce the complexity of the proof, we consider the case without literals (i.e. when no literal occurs in $S$ and $r$, and all of the variables in the schema $S^G$ are in $S^{\Delta}$). This proof can be extended to include literals by considering the the post-processing of the mappings, as demonstrated in the proof of Theorem \ref{theorem:approachEquality}. 
More precisely, we reformulate Theorem \ref{newTheoremSimple} as follows:

\begin{lem} \label{lemmaOldTheorem} For all rules $r: A \rightarrow C$ and for all triplestore schemas $S$ such that $S^\Delta = vars(S^G)$ and that no literal occurs in $r$ and $S^G$:

$\Rightarrow$) for all triple patterns $\Nto \in r(S) \setminus S$, 
for all triples $\Ntoc$ in an instance of \Nto, 
there exists a triple pattern $\NtoPrime \in \score(S,r) \setminus S$ 
s.t. $\{\Ntoc\} \in \mathds{I}(\{\NtoPrime\})$

$\Leftarrow$) for all triple patterns $\NtoPrime \in \score(S,r) \setminus S$, for all triples $\Ntoc$ in an instance of  $\NtoPrime$ there exists a triple pattern $\Nto \in r(S) \setminus S$ s.t. $\{\Ntoc\} \in \mathds{I}(\{\Nto\})$.

\end{lem}

\begin{Proof}
$\Rightarrow$) Given the premises of this part of Lemma \ref{lemmaOldTheorem}, there must exist a graph $I \in \mathds{I}(\{S\}$ on which an application of rule $r$ generates $\Ntoc$. For this to happen, there must exist a mapping $m \in \llbracket A\rrbracket_{I}$, such that $\Ntoc \in m(C)$. Obviously, the set of triples on which $A$ matches into $I$ via $m$ is $m(A) \subseteq I$.

For all triples $\Nta \in m(A)$ there exists a triple pattern $\Ntao \in S^G$ that models $\Nta$ (e.g.\ $\{\Nta\} \in \mathds{I}(\Ntao)$ ). We choose one of such triple pattern $\Ntao$ (in case multiple ones exist) and we call it the \emph{modelling triple} of $\Nta$. To prove the first part of Lemma \ref{lemmaOldTheorem} we will make use of the following proposition:

\emph{There exists a mapping $m^*$ and a rewriting $q \in \mathds{Q}(A)$, such that $m^*(q) \subseteq \mathds{S}(S)$ (thus $m^* \in \llbracket q\rrbracket_{\mathds{S}(S)}$), and that $m$ and $m^*$ agree on all variables except for those that are mapped to $\urilam$ in $m^*$. Formally, $dom(m) = dom(m^*)$ and for every variable $?v$ in $dom(m^*)$, either $m^*(?v) = m(?v)$, or $m^*(?v) = \urilam$.}

By proving the proposition above we know that our algorithm would extend the original schema graph $S^G$ with graph pattern $unpack(m^*(C))$, and the set of variables $S^\Delta$ with $vars(unpack(m^*(C)))$, where $unpack$ is a function that substitutes each $:\lambda$ in a graph with a fresh variable. If we take a triple pattern $c \in C$ such that $\Ntoc = m(\{c\})$, then $unpack(m^*(\{c\}))$ belongs to $\score(S,r)^G \setminus S^G$ (because $m^*$ matches a rewriting of the antecedent of the rule $r$ on the sandbox graph). Graph $unpack(m^*(\{c\}))$ also models triple $\Ntoc$. This is true because for each of the three positions $i$ of the schema triple pattern $c$, either $c[i]$ is a constant, and therefore this constant will also appear in position $i$ in both $unpack(m^*(\{c\}))[i]$ and $\Ntoc[i]$ or it is a variable, and by our proposition above either (a) $m^*(c)[i] = m(c)[i] = \Ntoc[i]$ and so $unpack(m^*(\{c\}))[i] = \Ntoc[i]$, or (b) $m^*(c)[i] = \urilam$ and therefore $unpack(m^*(\{c\}))[i]$ is a variable. We can then trivially recreate $\Ntoc$ as an instance of $unpack(m^*(\{c\}))$ by assigning to each variable $?v$ in $unpack(m^*(\{c\}))[i]$ the value $\Ntoc[i]$. Thus our algorithm would generate a triple pattern that models $\Ntoc$ and that would complete the proof of the first part ($\Rightarrow$) of the Theorem.

The proof of our proposition is as follows. Consider every position $i$ in every triple $\Nta \in m(A)$, and their corresponding modelling triple $\Ntao \in S$. Let $\NtA$ be the triple in $A$ such that $m(\NtA) = \Nta$. Let $\Nts$ be $\mathds{S}(\Ntao)$, and thus $\Nts \in \mathds{S}(S)$. We are now going to construct a query $q$, rewriting of $\mathds{Q}(A)$, such that $\llbracket q\rrbracket_{\mathds{S}(S)}$ gives us the mapping $m^*$ in our proposition. 
And to do this, we have to construct a triple pattern $\Ntq$ which will be the rewriting of $\NtA$ in $q$. By definition of how our rewritings are constructed, every element $\Ntq[i]$ is either $\NtA[i]$ or $\urilam$.

To set the value of each element in the triple pattern $\Ntq$ we consider the four possible cases that arise from $\NtA[i]$ and $\Ntao[i]$ being either a constant or a variable. In parallel, we consider the element-by-element evaluation of $q$ on $\mathds{S}(S)$ which generates $m^*$. 
\begin{enumerate}
    \item If $\NtA[i]$ and $\Ntao[i]$ are both constants, then since $\Nta \in m(A)$, it must be true that $\NtA[i] = \Nta[i]$. Moreover, since $\Ntao$ must be a model of $\Nta$, it follows that $\Ntao[i] = \Nta[i]$ and therefore $\Nts[i] = \Nta[i]$. We set element $\Ntq[i]$ to be the constant $\NtA[i]$, which matches the constant $\Nts[i]$.
    \item If $\NtA[i]$ is a constant but $\Ntao[i]$ is a variable, then we know that $\Nts[i] = \urilam$. We set element $\Ntq[i]$ to be the constant $\urilam$, which matches the constant $\Nts[i]$.
    \item If $\NtA[i]$ is a variable $?v$ and $\Ntao[i]$ is a constant $x$, then we know that $\Nta[i] = \Nts[i] = x$. Therefore mapping $m$ must contain binding $?v \to x$. We set element $\Ntq[i]$ to be the variable $?v$, so that when we complete the construction of $q$, $m^*$ will contain mapping $?v \to x$.
    \item If $\NtA[i]$ and $\Ntao[i]$ are both variables, it must be that $\Nts[i] = \urilam$.
    If it exists a triple $\NtAPrime \in A$ and a modelling triple $\NtAoPrime$ of $m(\NtAPrime)$, and position $j$ such that $\NtAPrime[i]$ is variable $?v$ and $\NtAoPrime[i]$ is a constant, then we set element $\Ntq[i]$ to be the constant $\urilam$ ($\Nts[i]$ will be $\urilam$ in our sandobox graph). Note that even though we don't use variable $\NtAPrime[i]$ in this part of our rewriting, the aforementioned existence of triple $\NtAPrime$ will force $m^*$ to contain a binding for this variable, and this binding will be in $m$ because this pair $\NtAPrime[i]$ and  $\NtAoPrime[i]$ will be treated by case $3$.
     Otherwise we set element $\Ntq[i]$ to be the variable $\NtA[i]$, which will bind with the value $\urilam$ of $\Nts[i]$ generating binding $?v \to \urilam$. 
\end{enumerate}
This proves our proposition, as we have constructed a query $q$, which belongs to $\mathds{Q}(A)$, which matches $\mathds{S}(S)$ and generates a mapping $m*$ which agrees on $m$ on all variables, except for those which $m^*$ maps to $\urilam$. 

$\Leftarrow$) 
Given the premises of the second direction of Lemma \ref{lemmaOldTheorem}, there must exist a mapping $m^*$ and a query $q \in \mathds{Q}(A)$ such that: $\NtoPrime \in unpack(m^*(C))$, $m^*(q) \subseteq \mathds{S}(S)$ and $dom(m^*) = dom(A)$. Note that $\mathds{S}(S)$ is an instance of $S$. There must exist a triple pattern $\Ntc$ in $C$ such that $unpack(m^*(\{\Ntc\})) = \NtoPrime$.

Note that $unpack$ transforms a triple  in a triple pattern by changing occurrences of $\lambda$ into fresh variables, if no $\lambda$ occurs in the triple, the triple remains unchanged.

For all positions $i$ of $\Ntc$, if $\Ntc[i]$ is a variable $?v$, given that $unpack(m^*(\{\Ntc\}))$ models $\Ntoc$, then mapping $m^*$ either contains the binding $?v \to \Ntoc[i]$, or it contains $?v \to \urilam$. Now consider the mapping $m$ generated by modifying mapping $m^*$ in the following way. For each position $i$ of $\Ntc$, if $\Ntc[i]$ is a variable $?v$ and $?v \to \urilam \in m^*$, substitute this binding with $?v \to \Ntoc[i]$. It is trivial to see that $\Ntoc \in m(C)$. We will now show that this mapping $m$ can be computed by evaluating the original query $A$ on a specific instance of $S$ by showing that $m(A)$ is an instance of $S$. By virtue of how query $q$ is constructed, we know that all of its elements in all of its triples must be the same as $A$, or the constant $\urilam$ instead.

As an intermediate step, we show that $m(q)$ is an instance of $S$. Since we know that $m^*(q)$ is an instance of $S$, and that $\urilam$ cannot occur in the triples in $S$, every occurrence of $\urilam$ in $m^*(q)$ must be generated from a variable $?v$ in a triple in $S$. Since each variable in $S^G$ occurs only once, we can change any occurrence of $\urilam$ in $m^*(q)$ into any other constant and still have an instance of $S$, since, intuitively, in the corresponding positions $S$ contains different variables. Given that the difference between $m$ and $m^*$ is that for some variables that $m^*$ binds to $\urilam$, $m$ binds them to a different constant, it is easy to see that the $m(q)$ can be generated from $m^*(q)$ only by changing some occurrences of $\lambda$ in its triples into a different constant. Thus $m(q)$ is an instance of $S$.

Now we are going to show that $m(A)$ is an instance of $S$. For all triples $\NtAtwo \in A$, our query rewriting algorithm has produced a $\Nta \in a$ for which, for each position $i$, we know that $\Nta[i] = \NtAtwo[i]$ or $\Nta[i] = \urilam$. 
Obviously for $m(\Nta) \in m(q)$, the same holds, i.e., for every position $i$, $m(\Nta)[i] = m(\NtAtwo)[i]$ or $m(\Nta)[i] = \urilam$.  
Since $m(\Nta)$ is an instance of $S$, there must exist a mapping $\gamma$ and a triple pattern $\Nts \in S^G$ such that 
$\gamma(\{\Nts\}) = m(\Nta)$. Since $\Nts$ cannot contain $\urilam$, for every position $i$ where $m(\Nta)[i] = \urilam$, $\Nts[i]$ contains a variable $?v_i$ which does not occur anywhere else in $S^G$ (by virtue of variables occurring only once in the schema); we can obtain the mapping from $\Nts$ to $m(\NtAtwo)$ by substituting each such binding $?v_i \to \urilam$ with $?v_i \to m(\NtAtwo)[i]$ in $\gamma$. For all triples $\NtAtwo \in A$, $m(\NtAtwo)$ is an instance of $S$, and therefore $\bigcup_{\NtAtwo \in A} \{m(\NtAtwo)\}$, which equals $m(A)$, is an instance of $S$.
\end{Proof}

\onecolumn
\section{Extended Example of Critical Instance}
Here is the complete listing of the critical instance of schema $S_1$, and rule $r_2$ from our example. 

\footnotesize \begin{lstlisting}

$\mathds{C}(S_1,r_2)$ = $\{$
 $\gpattern{\urilam}{\texttt{sosa:observedProperty}}{\uri{CO\_Danger}}$,
 $\gpattern{\texttt{sosa:observedProperty}}{\texttt{sosa:observedProperty}}{\uri{CO\_Danger}}$
 $\gpattern{\uri{CO\_Danger}}{\texttt{sosa:observedProperty}}{\uri{CO\_Danger}}$
 $\gpattern{\uri{WorkerTag}}{\texttt{sosa:observedProperty}}{\uri{CO\_Danger}}$
 $\gpattern{\texttt{sosa:hasFeatureOfInterest}}{\texttt{sosa:observedProperty}}{\uri{CO\_Danger}}$
 $\gpattern{\uri{TunnelA}}{\texttt{sosa:observedProperty}}{\uri{CO\_Danger}}$
 $\gpattern{\texttt{sosa:hasResult}}{\texttt{sosa:observedProperty}}{\uri{CO\_Danger}}$
 
 $\gpattern{\urilam}{\texttt{sosa:observedProperty}}{\uri{WorkerTag}}$,
 $\gpattern{\texttt{sosa:observedProperty}}{\texttt{sosa:observedProperty}}{\uri{WorkerTag}}$,
 $\gpattern{\uri{CO\_Danger}}{\texttt{sosa:observedProperty}}{\uri{WorkerTag}}$,
 $\gpattern{\uri{WorkerTag}}{\texttt{sosa:observedProperty}}{\uri{WorkerTag}}$,
 $\gpattern{\texttt{sosa:hasFeatureOfInterest}}{\texttt{sosa:observedProperty}}{\uri{WorkerTag}}$,
 $\gpattern{\uri{TunnelA}}{\texttt{sosa:observedProperty}}{\uri{WorkerTag}}$,
 $\gpattern{\texttt{sosa:hasResult}}{\texttt{sosa:observedProperty}}{\uri{WorkerTag}}$,
 
 $\gpattern{\urilam}{\texttt{sosa:hasFeatureOfInterest}}{\uri{TunnelA}}$,
 $\gpattern{\texttt{sosa:observedProperty}}{\texttt{sosa:hasFeatureOfInterest}}{\uri{TunnelA}}$,
 $\gpattern{\uri{CO\_Danger}}{\texttt{sosa:hasFeatureOfInterest}}{\uri{TunnelA}}$,
 $\gpattern{\uri{WorkerTag}}{\texttt{sosa:hasFeatureOfInterest}}{\uri{TunnelA}}$,
 $\gpattern{\texttt{sosa:hasFeatureOfInterest}}{\texttt{sosa:hasFeatureOfInterest}}{\uri{TunnelA}}$,
 $\gpattern{\uri{TunnelA}}{\texttt{sosa:hasFeatureOfInterest}}{\uri{TunnelA}}$,
 $\gpattern{\texttt{sosa:hasResult}}{\texttt{sosa:hasFeatureOfInterest}}{\uri{TunnelA}}$,
 
 $\gpattern{\urilam}{\texttt{sosa:hasResult}}{\urilam}$
 $\gpattern{\texttt{sosa:observedProperty}}{\texttt{sosa:hasResult}}{\urilam}$
 $\gpattern{\texttt{sosa:observedProperty}}{\texttt{sosa:hasResult}}{\texttt{sosa:observedProperty}}$
 $\gpattern{\texttt{sosa:observedProperty}}{\texttt{sosa:hasResult}}{\uri{CO\_Danger}}$
 $\gpattern{\texttt{sosa:observedProperty}}{\texttt{sosa:hasResult}}{\uri{WorkerTag}}$
 $\gpattern{\texttt{sosa:observedProperty}}{\texttt{sosa:hasResult}}{\texttt{sosa:hasFeatureOfInterest}}$
 $\gpattern{\texttt{sosa:observedProperty}}{\texttt{sosa:hasResult}}{\uri{TunnelA}}$
 $\gpattern{\texttt{sosa:observedProperty}}{\texttt{sosa:hasResult}}{\texttt{sosa:hasResult}}$
 $\gpattern{\texttt{sosa:observedProperty}}{\texttt{sosa:hasResult}}{\lit{1}}$
 
 $\gpattern{\uri{CO\_Danger}}{\texttt{sosa:hasResult}}{\urilam}$
 $\gpattern{\uri{CO\_Danger}}{\texttt{sosa:hasResult}}{\texttt{sosa:observedProperty}}$
 $\gpattern{\uri{CO\_Danger}}{\texttt{sosa:hasResult}}{\uri{CO\_Danger}}$
 $\gpattern{\uri{CO\_Danger}}{\texttt{sosa:hasResult}}{\uri{WorkerTag}}$
 $\gpattern{\uri{CO\_Danger}}{\texttt{sosa:hasResult}}{\texttt{sosa:hasFeatureOfInterest}}$
 $\gpattern{\uri{CO\_Danger}}{\texttt{sosa:hasResult}}{\uri{TunnelA}}$
 $\gpattern{\uri{CO\_Danger}}{\texttt{sosa:hasResult}}{\texttt{sosa:hasResult}}$
 $\gpattern{\uri{CO\_Danger}}{\texttt{sosa:hasResult}}{\lit{1}}$
 
 $\gpattern{\uri{WorkerTag}}{\texttt{sosa:hasResult}}{\urilam}$
 $\gpattern{\uri{WorkerTag}}{\texttt{sosa:hasResult}}{\texttt{sosa:observedProperty}}$
 $\gpattern{\uri{WorkerTag}}{\texttt{sosa:hasResult}}{\uri{CO\_Danger}}$
 $\gpattern{\uri{WorkerTag}}{\texttt{sosa:hasResult}}{\uri{WorkerTag}}$
 $\gpattern{\uri{WorkerTag}}{\texttt{sosa:hasResult}}{\texttt{sosa:hasFeatureOfInterest}}$
 $\gpattern{\uri{WorkerTag}}{\texttt{sosa:hasResult}}{\uri{TunnelA}}$
 $\gpattern{\uri{WorkerTag}}{\texttt{sosa:hasResult}}{\texttt{sosa:hasResult}}$
 $\gpattern{\uri{WorkerTag}}{\texttt{sosa:hasResult}}{\lit{1}}$
 
 $\gpattern{\texttt{sosa:hasFeatureOfInterest}}{\texttt{sosa:hasResult}}{\urilam}$
 $\gpattern{\texttt{sosa:hasFeatureOfInterest}}{\texttt{sosa:hasResult}}{\texttt{sosa:observedProperty}}$
 $\gpattern{\texttt{sosa:hasFeatureOfInterest}}{\texttt{sosa:hasResult}}{\uri{CO\_Danger}}$
 $\gpattern{\texttt{sosa:hasFeatureOfInterest}}{\texttt{sosa:hasResult}}{\uri{WorkerTag}}$
 $\gpattern{\texttt{sosa:hasFeatureOfInterest}}{\texttt{sosa:hasResult}}{\texttt{sosa:hasFeatureOfInterest}}$
 $\gpattern{\texttt{sosa:hasFeatureOfInterest}}{\texttt{sosa:hasResult}}{\uri{TunnelA}}$
 $\gpattern{\texttt{sosa:hasFeatureOfInterest}}{\texttt{sosa:hasResult}}{\texttt{sosa:hasResult}}$
 $\gpattern{\texttt{sosa:hasFeatureOfInterest}}{\texttt{sosa:hasResult}}{\lit{1}}$
 
 $\gpattern{\uri{TunnelA}}{\texttt{sosa:hasResult}}{\urilam}$
 $\gpattern{\uri{TunnelA}}{\texttt{sosa:hasResult}}{\texttt{sosa:observedProperty}}$
 $\gpattern{\uri{TunnelA}}{\texttt{sosa:hasResult}}{\uri{CO\_Danger}}$
 $\gpattern{\uri{TunnelA}}{\texttt{sosa:hasResult}}{\uri{WorkerTag}}$
 $\gpattern{\uri{TunnelA}}{\texttt{sosa:hasResult}}{\texttt{sosa:hasFeatureOfInterest}}$
 $\gpattern{\uri{TunnelA}}{\texttt{sosa:hasResult}}{\uri{TunnelA}}$
 $\gpattern{\uri{TunnelA}}{\texttt{sosa:hasResult}}{\texttt{sosa:hasResult}}$
 $\gpattern{\uri{TunnelA}}{\texttt{sosa:hasResult}}{\lit{1}}$
 
 $\gpattern{\texttt{sosa:hasResult}}{\texttt{sosa:hasResult}}{\urilam}$
 $\gpattern{\texttt{sosa:hasResult}}{\texttt{sosa:hasResult}}{\texttt{sosa:observedProperty}}$
 $\gpattern{\texttt{sosa:hasResult}}{\texttt{sosa:hasResult}}{\uri{CO\_Danger}}$
 $\gpattern{\texttt{sosa:hasResult}}{\texttt{sosa:hasResult}}{\uri{WorkerTag}}$
 $\gpattern{\texttt{sosa:hasResult}}{\texttt{sosa:hasResult}}{\texttt{sosa:hasFeatureOfInterest}}$
 $\gpattern{\texttt{sosa:hasResult}}{\texttt{sosa:hasResult}}{\uri{TunnelA}}$
 $\gpattern{\texttt{sosa:hasResult}}{\texttt{sosa:hasResult}}{\texttt{sosa:hasResult}}$
 $\gpattern{\texttt{sosa:hasResult}}{\texttt{sosa:hasResult}}{\lit{1}}$
  
 $\gpattern{\urilam}{\texttt{sosa:hasResult}}{\texttt{sosa:observedProperty}}$
 $\gpattern{\urilam}{\texttt{sosa:hasResult}}{\uri{CO\_Danger}}$
 $\gpattern{\urilam}{\texttt{sosa:hasResult}}{\uri{WorkerTag}}$
 $\gpattern{\urilam}{\texttt{sosa:hasResult}}{\texttt{sosa:hasFeatureOfInterest}}$
 $\gpattern{\urilam}{\texttt{sosa:hasResult}}{\uri{TunnelA}}$
 $\gpattern{\urilam}{\texttt{sosa:hasResult}}{\texttt{sosa:hasResult}}$
 $\gpattern{\urilam}{\texttt{sosa:hasResult}}{\lit{1}}$
$\}$
\end{lstlisting} 

\noindent The critical instance of $\mathds{C}(S_1,r_1)$ would be of similar length, and it would feature the \texttt{URI}   \uri{OffLimitArea} instead of the literal \lit{1}.

\end{document}